\documentclass[11pt]{article}

\usepackage{epsfig}

\begin{document}

\title{\LARGE An approximate solution for solar and supernova neutrino oscillation in matter}

\author{Rui Luo$^{1}$\thanks{Email: ruiluo@pku.edu.cn}\\
\\
$^{1}$ School of Physics, Peking University,\\
Beijing 100871, P. R. China}
\date{}

\maketitle

\begin{abstract}
By using Laplace transformation we developed an approximate
solution to describe neutrino oscillation probabilities in
arbitrary density matter. We show that this approximation solution
is valid when matter potential V satisfy $V< \Delta m^2/2E$ and
$\int VL<1$, where $L$ is the length of the neutrino oscillation .
Thus, the formula is useful for propagation of the solar or
supernova neutrinos with terrestrial matter effect.
\end{abstract}

\newpage

\section{Introduction}

By now, we have solar, supernova, atmospheric, reactor and
accelerator neutrino experiments to determine the mass splittings
and flavor mixings in neutrinos. In order to calculate the
survival and conversion probabilities of neutrinos passing through
the Earth, the matter effect \cite{Wolf:78} must be taken into
account. Previous work on this subject inculdes exact and
approximate expressions in the case of constant matter density
\cite{Barger:80}-\cite{Ble:04}, linear density \cite{Leh:00} and
exponential density \cite{Osl:99}. As for the case of arbitrary
matter density, approximate solutions have also be presented by
Akhmedov \cite{Akh:04a} , Peres \cite{Per:99} and Ioannisian
\cite{Ioan:04},\cite{Ioan:04a}.

Motivated by Ioannisian's paper\cite{Ioan:04}, we derived an
approximate solution to MSW equation by using \textit{Laplace
Transformation}. The form of the formula in this note is so simple
and transparent that it is no difficult to calculate up to the
high order terms. So, with this formula, we could simplify the
numerical calculation considerably. We also show that the solution
is valid for the case of low energy neutrino such as solar and
supernova neutrinos, and the approximation is effective when the
baseline length is not too long.

The outline of the paper is as follows. First, we investigate the
general approximate solution for electron neutrino survival
probability with 2 flavors. Then we compare our formulas with
exact numerical results in some special case: uniform matter
density and linear matter density. And finally we discuss some
related issues.

\section{General approximate formula by using Laplace Transformation}
We consider the case of two-flavor (electron flavor $\nu_e$  and
the effective flavor $\nu_x$ - a linear combination of $\nu_{\mu}$
and $\nu_{\tau}$) neutrino oscillation for simplicity. The
neutrino mixing between flavor-eigenstates and mass-eigenstats
could be expressed as
\begin{equation}\label{Leptmixing}
\nu_f(x)=U(\theta)\nu_m(x),
\end{equation}
with $ \nu_f=(\nu_e(x), \nu_x(x))^T$ and $\nu_m=(\nu_1(x),
\nu_2(x))^T$ are the flavor eigenstates and mass eigenstates
respectively. And the lepton mixing matrix is given by
\begin{equation}\label{mixingmatrix}
U(\theta)= \left( \begin{tabular}{cc}
$\cos \theta$ & $\sin \theta$ \\
$ - \sin \theta$ & $\cos \theta$
\end{tabular}
\right).
\end{equation}
In order to find the neutrino oscillation probabilities in matter,
we have to solve the Schr\"{o}dinger equation for flavor
eigenstates
\begin{equation}\label{Schrodinger}
i \frac{d}{dx}\left( \begin{tabular}{c}
$\nu_e(x)$  \\
$\nu_x(x)$
\end{tabular}
\right) = H(x) \left( \begin{tabular}{c}
$\nu_e(x)$  \\
$\nu_x(x)$
\end{tabular}
\right).
\end{equation}
And the effective Hamiltonian for propagation of neutrinos in
matter
\begin{equation}\label{Hamiltonian}
 H(t)=\frac{1}{2E} \left[ U\left( \begin{tabular}{cc}
$m_1^2$& \\
 & $m_2^2$
\end{tabular}
\right)U^\dagger + \left( \begin{tabular}{cc}
$A(x)$& \\
 & $0$
\end{tabular}
\right)\right],
\end{equation}
where $A(x)=2\sqrt{2}G_FN_e(x)E$ is the effective potential term.
With $G_F$ is the Fermi constant, $N_e(x)$ stands for the electron
density in matter at point $x$, and $E$ is the neutrino beam
energy. We take
\begin{equation}\label{Anum} A=3.8\times 10^{-4}eV^2 \left(
\frac{Y_e\rho}{2.5 g/cm^3} \right) \left( \frac{E}{1~GeV} \right
).
\end{equation}
Notice that the sign of the matter potential is positive
for neutrinos and negative for anti-neutrinos.

Applying Laplace transformation to the Schr\"{o}dinger equation,
we get

\begin{eqnarray}\label{Laplace}
& &i s\left( \begin{tabular}{c}
${\cal L} [ \nu_e(x) ]$  \\
${\cal L} [ \nu_x(x) ]$
\end{tabular}
\right)-i \left( \begin{tabular}{c}
$ \nu_e(0) $  \\
$ \nu_x(0) $
\end{tabular}
\right) \nonumber \\
 &  =& \left( \begin{tabular}{cc}
$\Delta_1\cos^2\theta+\Delta_2\sin^2\theta$ & $-\Delta_1\sin\theta\cos\theta+\Delta_2\sin\theta\cos\theta$ \\
$-\Delta_1\sin\theta\cos\theta+\Delta_2\sin\theta\cos\theta$ &
$\Delta_1\sin^2\theta+\Delta_2\cos^2\theta$
\end{tabular}
\right) \left( \begin{tabular}{c}
${\cal L} [ \nu_e(x) ]$  \\
${\cal L} [ \nu_x(x) ]$
\end{tabular}
\right) \nonumber \\
 &&  +\left( \begin{tabular}{c}
${\cal L}[V(x) \nu_e(x) ]$  \\
$0$
\end{tabular}
\right).
\end{eqnarray}

Here we define $\Delta_1=m_1^2/2E$, $\Delta_2=m_2^2/2E$ and
$V(x)=A(x)/2E$. To find the relation between $\nu_e(x)$ and
$V(x)$, we rewrite this equation as
\begin{eqnarray}\label{Laplace1}
\left( \begin{tabular}{c}
${\cal L} [ \nu_e(x) ]$  \\
${\cal L} [ \nu_x(x) ]$
\end{tabular}
\right) &=& \left( \begin{tabular}{cc}
$is-\Delta_1\cos^2\theta-\Delta_2\sin^2\theta$ & $\Delta_1\sin\theta\cos\theta-\Delta_2\sin\theta\cos\theta$ \\
$\Delta_1\sin\theta\cos\theta-\Delta_2\sin\theta\cos\theta$ &
$is-\Delta_1\sin^2\theta-\Delta_2\cos^2\theta$
\end{tabular}
\right)^{-1}  \nonumber \\
 & & \times\left[ i\left(\begin{tabular}{c}
$\nu_e(0)$  \\
$\nu_x(0)$
\end{tabular}\right)+\left(\begin{tabular}{c}
${\cal L}[V(x) \nu_e(x) ]$  \\
$0$
\end{tabular}\right)
\right].
\end{eqnarray}
It is straightforward to set the initial condition as $\nu_e(0)=1$
and $\nu_x(0)=0$ when we calculate the electron neutrino survival
probability $P_{ee}$ and conversion probability $P_{ex}$. Thus Eq.
(\ref{Laplace}) could be expressed explicitly as
\begin{eqnarray}\label{equ1}
 {\cal L} [ \nu_e(x)
 ]&=&\frac{\cos^2\theta}{(s+i\Delta_1)}+\frac{\sin^2\theta}{(s+i\Delta_2)}\nonumber \\
 &&-i\left(\frac{\cos^2\theta}{(s+i\Delta_1)}+\frac{\sin^2\theta}{(s+i\Delta_2)}\right){\cal L}[V(x) \nu_e(x)
 ],
\end{eqnarray}
and
\begin{eqnarray}\label{equ2}
 {\cal L} [ \nu_x(x)
 ]&=&\sin\theta\cos\theta\left( \frac{1}{s+i\Delta_1}-\frac{1}{s+i\Delta_1}\right)\nonumber \\
 &&-i\sin\theta\cos\theta\left( \frac{1}{s+i\Delta_1}-\frac{1}{s+i\Delta_1}\right){\cal L}[V(x) \nu_e(x)
 ].
\end{eqnarray}

Then we apply Inverse Laplace Transformation to Eq. (\ref{equ1})
and (\ref{equ2}), and we arrive at two integral equations of the
oscillation amplitude of $\nu_e$ and $\nu_x$:
\begin{eqnarray}\label{inteq1}
  \nu_e(x)
 &=&\cos^2\theta e^{-i\Delta_1x}+\sin^2\theta e^{-i\Delta_2x} \nonumber \\
&& -i\int_0^x dy\left(\cos^2\theta
e^{-i\Delta_1(x-y)}+\sin^2\theta e^{-i\Delta_2(x-y)}\right)V(y)
\nu_e(y),
\end{eqnarray}
\begin{eqnarray}\label{inteq2}
  \nu_x(x)
 &=&\sin\theta\cos\theta \left(e^{-i\Delta_1x}- e^{-i\Delta_2x} \right)\nonumber \\
&& -i\sin\theta\cos\theta\int_0^x dy\left(
e^{-i\Delta_1(x-y)}-e^{-i\Delta_2(x-y)}\right)V(y) \nu_e(y).
\end{eqnarray}
Notice that the third term in the right side of Eq. (\ref{inteq1})
and (\ref{inteq2}) is a convolution integral. We will deal with
Eq. (\ref{inteq1}) first, and it is clear that Eq. (\ref{inteq2})
could
be solved straightforwardly with the result of Eq. (\ref{inteq1}). \\
In order to get the approximate solution to this equation, it is
convenient to define an operator
\begin{eqnarray}\label{defop}
 K(\nu_e(x))&=&-i\int_0^x dy\left(\cos^2\theta
e^{-i\Delta_1(x-y)}+\sin^2\theta e^{-i\Delta_2(x-y)}\right)V(y)
\nu_e(y)\nonumber \\&=& -i\cos^2\theta e^{-i\Delta_1x}\int_0^x dy
e^{i\Delta_1y}V(y) \nu_e(y)\nonumber \\&&-i\sin^2\theta
e^{-i\Delta_2y}\int_0^x dy e^{i\Delta_2y}V(y) \nu_e(y).
\end{eqnarray}
Thus, Eq. (\ref{defop}) could be rewrited as
\begin{eqnarray}\label{rewt}
\nu_e(x)=\cos^2\theta e^{-i\Delta_1x}+\sin^2\theta e^{-i\Delta_2x}
+K(\nu_e(x)).
\end{eqnarray}
Thus, if $\parallel K(\nu_e(x))\parallel<1$, the solution of
$\nu_e(x)$ could be expressed in a series expansion form:
\begin{eqnarray}\label{appsol}
\nu_e(x)=\left(1+K+K^2+K^3+\cdots \right)\left(\cos^2\theta
e^{-i\Delta_1x}+\sin^2\theta e^{-i\Delta_2x} \right).
\end{eqnarray}
Inserting Eq. (\ref{rewt}) into Eq. (\ref{inteq2}), the
oscillation amplitude of $\nu_x$
\begin{eqnarray}\label{xappsol}
  \nu_x(x)
 &=&\sin\theta\cos\theta \left(e^{-i\Delta_1x}- e^{-i\Delta_2x} \right)\nonumber \\
&& -i\sin\theta\cos\theta\int_0^x dy\left(
e^{-i\Delta_1(x-y)}-e^{-i\Delta_2(x-y)}\right)V(y) \nu_e(y).
\end{eqnarray}

A straightforward calculation leads to the $\nu_e$ survival
probability $P_{\nu_e \rightarrow \nu_e}=|\nu_e(x)|^2$ and
conversion probability $P_{\nu_e \rightarrow \nu_x}=|\nu_x(x)|^2$.
We note that other oscillation probabilities such as $P_{\nu_x
\rightarrow \nu_x}$ or $P_{\nu_x \rightarrow \nu_e}$ could be
obtained with the same method but different initial condition
($[\nu_e, \nu_x]^T=[0,1]^T$).

\section{Numerical discussion and testing of accuracy}

Eq. (\ref{appsol}) in the last section is a general formula. In
this section, we discuss the qualitative behavior of this formula.
First let us consider the case of constant matter density
($V(x)=V$). And with the result we can derive a raw applicability
condition for this formula.

Eq. (\ref{appsol}) could be expanded in orders explicitly as the
matter potential is a constant $V(x)=V$. The first three orders of
the amplitude of electron neutrino oscillation $\nu_e(x)$ are
\begin{eqnarray}\label{order0}
\nu_e(x)^{[0]}=\cos^2\theta e^{-i\Delta_1x}+\sin^2\theta
e^{-i\Delta_2x},
\end{eqnarray}

\begin{eqnarray}\label{order1}
\nu_e(x)^{[1]}&=&-i\cos^4\theta V x e^{-i\Delta_1x}-i\sin^4\theta
V x e^{-i\Delta_2x}\nonumber \\
&&-\sin^2\theta\cos^2\theta
\frac{2V}{\Delta_1-\Delta2}(e^{-i\Delta_2x}-e^{-i\Delta_1x}),
\end{eqnarray}
\begin{eqnarray}\label{order2}
\nu_e(x)^{[2]}&=&-\cos^6\theta \frac{V^2 x^2}{2}
e^{-i\Delta_1x}-\sin^6\theta \frac{V^2 x^2}{2}
e^{-i\Delta_2x}\nonumber \\
&&-3i\sin^2\theta\cos^4\theta \frac{V^2 x}{\Delta_1-\Delta_2}
e^{-i\Delta_1x}+3i\sin^4\theta\cos^2\theta \frac{V^2
x}{\Delta_1-\Delta_2} e^{-i\Delta_2x}\nonumber \\
&&+\frac{3}{4}\sin^22\theta\cos2\theta
\frac{V^2}{(\Delta_1-\Delta_2)^2}(e^{-i\Delta_2x}-e^{-i\Delta_1x}).
\end{eqnarray}
We can see that there are two expansion parameters in Eq.
(\ref{order0})-(\ref{order2}): $Vx$ ,which has appeared in
\cite{Akh:01} and $\frac{V}{\Delta_1-\Delta_2}$. If the number
density of the electrons $N_e=2.26$ \cite{Barger:00}, and
$x=D_{Earth}$, the diameter of the earth, we could find that
$Vx\approx 0.4$. Since in the case of earth-induced matter effect,
$x\leq D_{earth}$, we could estimate that $Vx<0.4$ thus it could
be proved as an effective expansion parameter. Another expansion
parameter $\frac{V}{\Delta_1-\Delta_2}$ is similar to the
$\epsilon(x)$ in Ioannisian's paper\cite{Ioan:04} and
\cite{Ioan:04a}, which would be small enough
($\frac{V}{\Delta_1-\Delta_2}<0.6$ if $E<50MeV$
\cite{Ioan:04a})for approximation in solar and supernova neutrinos
(low energy). \\

The oscillation probability up to the zero-th order (we set
$\Delta_1=0$):
\begin{eqnarray}\label{Porder0}
P^{[0]}_{e\rightarrow e}&=&\mid \nu_e(x)^{[0]}\mid^2 \nonumber\\
&=&1-\sin^22\theta\sin^2(\frac{\Delta_2}{2}x),
\end{eqnarray}
first order:
\begin{eqnarray}\label{Porder1}
P^{[1]}_{e\rightarrow e}&=&\mid \nu_e(x)^{[0]}+ \nu_e(x)^{[1]}\mid^2 \nonumber\\
&=& 1-\sin^22\theta\sin^2(\frac{\Delta_2}{2}x)\nonumber \\
&&-\frac{Vx}{2}\sin^22\theta\cos2\theta\sin\Delta_2x+
\frac{V}{\Delta_2}\sin^22\theta\cos2\theta(1-\cos\Delta_2x),\nonumber\\
\end{eqnarray}
second order:
\begin{eqnarray}\label{Porder2}
P^{[2]}_{e\rightarrow e}&=&\mid \nu_e(x)^{[0]}+ \nu_e(x)^{[1]}+\nu_e(x)^{[2]}\mid^2 \nonumber\\
&=& 1-\sin^22\theta\sin^2(\frac{\Delta_2}{2}x)\nonumber \\
&&-\frac{Vx}{2}\sin^22\theta\cos2\theta\sin\Delta_2x+
\frac{V}{\Delta_2}\sin^22\theta\cos2\theta(1-\cos\Delta_2x)\nonumber \\
&&-\frac{V^2x^2}{16}\sin^24\theta\cos\Delta_2x\nonumber\\
&&+\frac{V^2x}{\Delta_2}\sin^2\theta\cos^2\theta(4\sin^4\theta-12\sin^2\theta\cos^2\theta+4\cos^4\theta)\sin\Delta_2x\nonumber\\
&&+\frac{V^2}{\Delta_2^2}\sin^2\theta\cos^2\theta(6\sin^4\theta-20\sin^2\theta\cos^2\theta+6\cos^4\theta)(\cos\Delta_2x-1).\nonumber\\
\end{eqnarray}

Inserting Eq. (\ref{order0})-(\ref{order1}) to Eq. (\ref{inteq2}),
we get the amplitude of $\nu_x(x)$ up to the first order
\begin{eqnarray}\label{ex0}
\nu_x(x)^{[0]}=\sin\theta\cos\theta
\left(e^{-i\Delta_1x}-e^{-i\Delta_2x} \right),
\end{eqnarray}
and
\begin{eqnarray}\label{ex1}
\nu_x(x)^{[1]}&=&\sin\theta\cos\theta
\left(e^{-i\Delta_1x}-e^{-i\Delta_2x} \right)\nonumber\\
&&-i\sin\theta\cos^3\theta V x e^{-i\Delta_1x}
-i\sin^3\theta\cos\theta V x
e^{-i\Delta_2x})\nonumber\\
&&+\frac{1}{2}\sin2\theta\cos2\theta\frac{e^{-i\Delta_1x}-e^{-i\Delta_2x}
}{\Delta_1-\Delta_2}.
\end{eqnarray}

And from Eq. (\ref{ex1}), we get the oscillation probability (omit
the second order term)
\begin{eqnarray}\label{Pex1}
P^{[1]}_{e\rightarrow x}&=&\mid \nu_x(x)^{[1]}\mid^2 \nonumber\\
&=& \sin^22\theta\sin^2(\frac{\Delta_2}{2}x)\nonumber \\
&&+\frac{Vx}{2}\sin^22\theta\cos2\theta\sin\Delta_2x-
\frac{V}{\Delta_2}\sin^22\theta\cos2\theta(1-\cos\Delta_2x).\nonumber\\
\end{eqnarray}

We find $P^{[1]}_{e\rightarrow e}+P^{[1]}_{e\rightarrow x}=1$,
which could serve as a cross check of the formula.

Since in the case of constant matter density, the convergency
condition of this approximation is $Vx<1$ and
$\frac{V}{\Delta_1-\Delta_2}<1$, we can arrive at a raw
convergency condition for arbitrary density:
\begin{eqnarray}\label{cond1}
V_{max}x<1,
\end{eqnarray}
and
\begin{eqnarray}\label{cond2}
\frac{V_{max}}{\Delta_1-\Delta_2}<1.
\end{eqnarray}
It is transparent that both of the two conditions could be widely
satisfied if we investigate the terrestrial matter effect of low
energy neutrino.

\begin{figure}
\begin{center}
\epsfig{file=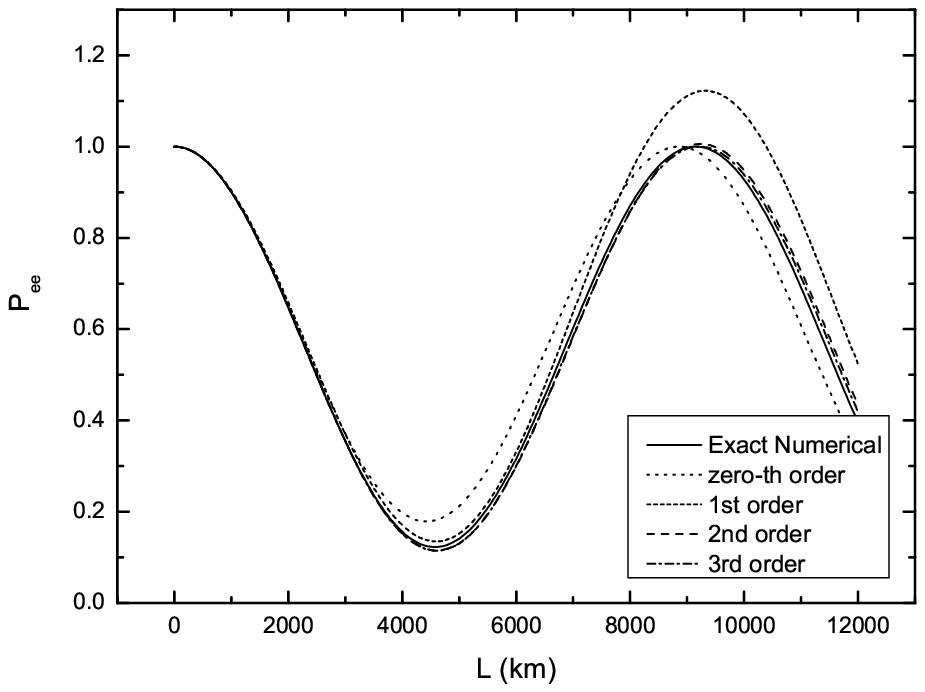,width=12cm} \caption{Electron neutrino
survival probability as a function of length of baseline $L$ with
$E=10MeV$ and $\rho=6.0 g/cm^3$ ($V \approx 2\times 10^{-13}eV$)
and best-fit KamLAND data \cite{KamLAND} parameter values: $\Delta
m_{sol}^2=7.1\times 10^{-5}eV^2$ and $\theta_{sol}=32.5^\circ$.}
\label{fig1}
\end{center}
\end{figure}

\begin{figure}
\begin{center}
\epsfig{file=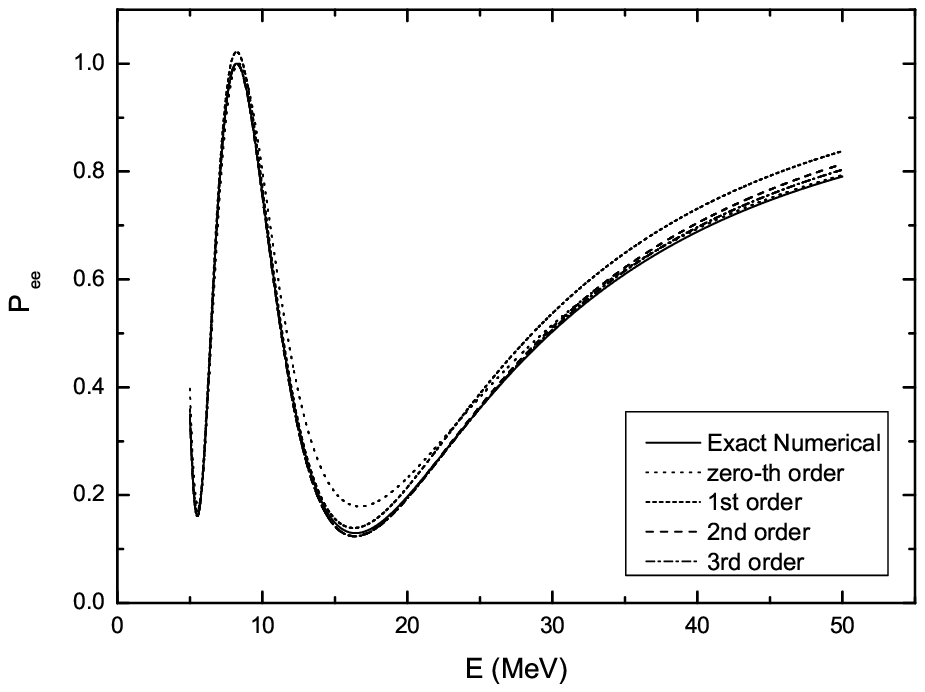,width=12cm} \caption{Electron neutrino
survival probability as a function of energy $E$ with baseline
length $L=7400~km$ and $\rho=6.0 g/cm^3$ ($V \approx 2\times
10^{-13}eV$) and best-fit KamLAND data \cite{KamLAND} parameter
values: $\Delta m_{sol}^2=7.1\times 10^{-5}eV^2$ and
$\theta_{sol}=32.5^\circ$.} \label{fig2}
\end{center}
\end{figure}

In order to examine the reliability of this formula, we compare
 the electron neutrino survival probability $P_{e\rightarrow e}$ obtained from
 the zeroth order correction, first order correction and second order correction
 with the exact numerical result. In Fig. (\ref{fig1}) and (\ref{fig2}), we observe that the solution to the
 electron neutrino survival probability $P_{e\rightarrow e}$ is a good
 approximation already when the length of neutrino propagation
 ~$L<6000~km$ and neutrino energy $E<20 MeV$. As for the case of
  ~$L>6000~km$ and $E>20 MeV$, we have to calculate the
 approximate probability up to the second or third order.\\

\begin{figure}
\begin{center}
\epsfig{file=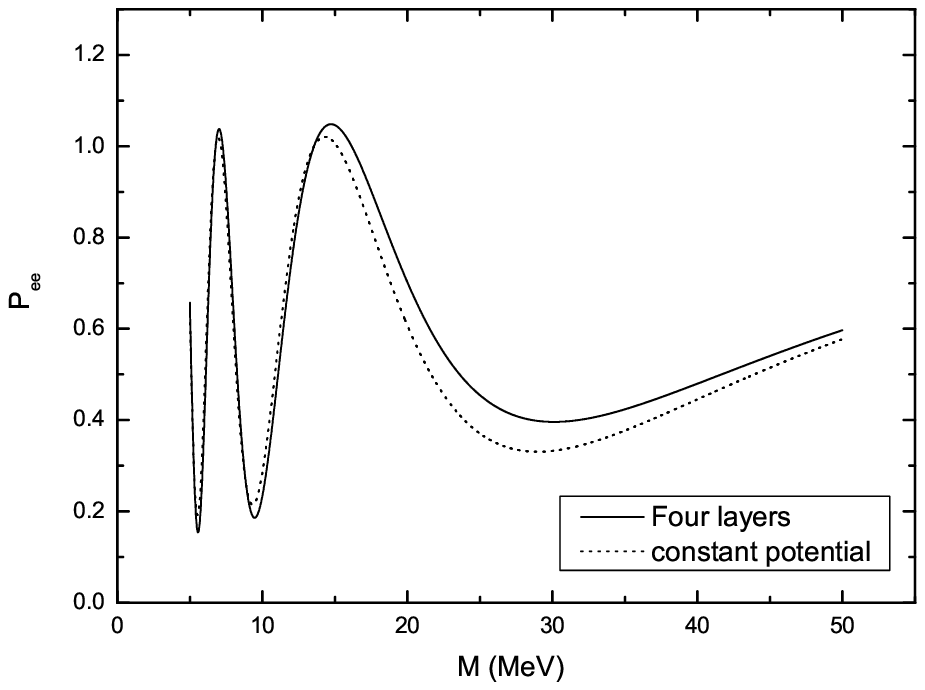,width=12cm} \caption{Neutrino oscillation
probability $P_{e\rightarrow e}$ vs neutrino energy, when the
neutrino flux pass through the center of the Earth with best-fit
KamLAND data \cite{KamLAND} parameter values: $\Delta
m_{sol}^2=7.1\times 10^{-5}eV^2$ and $\theta_{sol}=32.5^\circ$}
\label{earth}
\end{center}
\end{figure}

When the potential is a linear function $V(x)=a+bx$, the formula
is also useful. According to Eq. (\ref{appsol}) we can arrive at
an approximate solution up to the first order.
\begin{eqnarray}\label{lineorder0}
\nu_e(x)^{[0]}=\cos^2\theta e^{-i\Delta_1x}+\sin^2\theta
e^{-i\Delta_2x},
\end{eqnarray}
\begin{eqnarray}\label{lineorder1}
\nu_e(x)^{[1]}&=&-i\cos^4\theta a x e^{-i\Delta_1x}-i\sin^4\theta
ax e^{-i\Delta_2x}\nonumber \\
&&-\sin^2\theta\cos^2\theta
\frac{2a}{\Delta_1-\Delta_2}(e^{-i\Delta_2x}-e^{-i\Delta_1x})\nonumber \\
&&-i\cos^4\theta\frac{bx^2}{2}e^{-i\Delta_1x}-i\sin^4\theta\frac{bx^2}{2}e^{-i\Delta_2x}.\nonumber \\
\end{eqnarray}
So, the oscillation probability (we set $\Delta_1=0$)
\begin{eqnarray}\label{linePorder1}
P^{[1]}_{e\rightarrow e}&=&|\nu_e(x)^{[0]}+\nu_e(x)^{[1]}|^2  \nonumber \\
&\simeq& 1-\sin^22\theta\sin^2(\frac{\Delta_2}{2}x)\nonumber \\
&&-\frac{ax}{2}\sin^22\theta\cos2\theta\sin\Delta_2x+
\frac{a}{\Delta_2}\sin^22\theta\cos2\theta(1-\cos\Delta_2x)\nonumber \\
&&-\frac{bx^2}{4}\sin^22\theta\cos2\theta\sin\Delta_2x.
\end{eqnarray}
Here we calculate a special case: when the neutrino flux pass
through the center of the Earth, according to \cite{PREM}, the
potential could be expressed as four sections of linear functions
approximately.
\begin{eqnarray}
V(x)=a_1+bx ~~~~~~~~~~~0<x<x_1,\nonumber \\
V(x)=a_2+bx ~~~~~~~~~~x_1<x<x_2,\nonumber \\
V(x)=a_3-bx ~~~~~~~~~~x_2<x<x_3,\nonumber \\
V(x)=a_4-bx ~~~~~~~~~~x_3<x<x_4,
\end{eqnarray}
where $a_1=1.14\times10^{-13}[eV],~a_2=3.04\times10^{-13}[eV],~
a_3=6.84\times10^{-13}[eV],~a_4=4.94\times10^{-13}[eV],
~b=2.7\times10^{-17}[eV]/[km]$,
$x_1=2800km,~x_2=6200km,~x_3=9600km,~x_4=12000km$. With Eq.
(\ref{lineorder0}) and Eq. (\ref{lineorder1}), we obtain
\begin{eqnarray}
P_{e\rightarrow e}&=&
1-\sin^22\theta\sin^2(\frac{\Delta_2}{2}x)\nonumber\\
&&+2(\cos^2\theta+\sin^2\theta\cos\Delta_2x_4)
(f\sin^4\theta \sin\Delta_2x_4+\frac{g_1}{\Delta_2}\sin^2\theta\cos^2\theta)\nonumber\\
&&+2\sin^2\theta\sin\Delta_2x_4
(-f\cos^4\theta -f\sin^4\theta\cos\Delta_2x_4+\frac{g_2}{\Delta_2}\sin^2\theta\cos^2\theta),\nonumber\\
\end{eqnarray}
where
\begin{eqnarray}
f=a_1x_1+a_2x_2-a_2x_1+a_3x_3-a_3x_2+a_4x_4-a_4x_3+bx_2^2-\frac{bx_4^2}{2},
\end{eqnarray}
\begin{eqnarray}
g_1&=&a_1\cos\Delta_2(x_4-x_1)-a_1\cos\Delta_2x_4+a_2\cos\Delta_2(x_4-x_2)\nonumber\\
&&-a_2\cos\Delta_2(x_4-x_1)+a_3\cos\Delta_2(x_4-x_3)-a_3\cos\Delta_2(x_4-x_2)\nonumber\\
&&-a_4\cos\Delta_2(x_4-x_3)-a_1\cos\Delta_2x_1-a_2\cos\Delta_2x_2+a_2\cos\Delta_2x_1\nonumber\\
&&-a_3\cos\Delta_2x_3+a_3\cos\Delta_2x_2-a_4\cos\Delta_2x_4+a_4\cos\Delta_2x_3+a_1+a_4,\nonumber\\
\end{eqnarray}
\begin{eqnarray}
g_2&=&a_1\sin\Delta_2(x_4-x_1)-a_1\sin\Delta_2x_4+a_2\sin\Delta_2(x_4-x_2)\nonumber\\
&&-a_2\sin\Delta_2(x_4-x_1)+a_3\sin\Delta_2(x_4-x_3)-a_3\sin\Delta_2(x_4-x_2)\nonumber\\
&&-a_4\sin\Delta_2(x_4-x_3)-a_1\sin\Delta_2x_1-a_2\sin\Delta_2x_2+a_2\sin\Delta_2x_1\nonumber\\
&&-a_3\sin\Delta_2x_3+a_3\sin\Delta_2x_2-a_4\sin\Delta_2x_4+a_4\sin\Delta_2x_3,
\end{eqnarray}
The numerical result is showed in Fig.(\ref{earth}), compared with
the result of uniform density (we set the potential
$V=3\times10^{-13}[eV]$, which is the average density of the
Earth). And we can find that in such case, constant potential is
not a good approximation.

Furthermore, the formula of two neutrino species could be
 generalized in a straightforward way to the case of any neutrino
 species. If there are N types of neutrino involved, Eq. (\ref{equ1}) turns into
\begin{eqnarray}\label{equ1r}
 {\cal L} [ \nu_e(x)
 ]&=&\frac{a_1}{(s+i\Delta_1)}+\frac{a_2}{(s+i\Delta_2)}+~\cdots~+\frac{a_N}{(s+i\Delta_N)}\nonumber \\
 &&-i\left(\frac{a_1}{(s+i\Delta_1)}+\frac{a_2}{(s+i\Delta_2)}+~\cdots~+\frac{a_N}{(s+i\Delta_N)}\right)\nonumber \\
 &&\times {\cal L}[V(x) \nu_e(x) ],
\end{eqnarray}
where
\begin{eqnarray}
 a_i= U_{ei}U^*_{ei},
 \end{eqnarray}
and $U$ is the N-flavor mixing matrix. Thus, the solution could be
expressed as
\begin{eqnarray}\label{appsolr}
\nu_e(x)=\left(1+K'+K'^2+K'^3+\cdots \right)\left(a_1
e^{-i\Delta_1x}+a_2 e^{-i\Delta_2x}+\cdots+ a_N
e^{-i\Delta_Nx}\right).
\end{eqnarray}
Here the operator $K'$ is redefined as
\begin{eqnarray}\label{defopr}
 K'(\nu_e(x))&=&-i\int_0^x dy\left(a_1
e^{-i\Delta_1(x-y)}+a_2 e^{-i\Delta_2(x-y)}+\cdots+a_N
e^{-i\Delta_N(x-y)}\right)V(y) \nu_e(y).\nonumber\\
\end{eqnarray}

\section{Conclusion}
We have derived a series expansion formulation for neutrino
oscillation in arbitrary density matter. Examining the case of
constant density, we found that this formula is useful in
calculating electron survival and conversion probability with a
not too long baseline ( $L<12000~km$ ) and low energy (
$E<70~-~80~MeV$ ). As for some special cases, we can see that this
expansion up to second order is possible to reach a high accuracy.
We also show that this formulation could be extended to the case
of N-flavor neutrinos by adding similar extra terms to the
expansion operator $K$.

\section*{Acknowledgements}
The author would like to thank Professor Shou-Hua Zhu and Da-Xin
Zhang for helpful discussions.

\newpage

\begin{thebibliography}{99}


\bibitem{Wolf:78}
L.~Wolfenstein, ``Neutrino oscillation in matter,'' Phys. Rev.
{\bf D 17}, 2369 (1978).



\bibitem{Barger:80}
V.D.~Barger, K.~Whisnant, S.~Pakvasa and R.J.N.~Phillips, ``Matter
effcts on three-neutrino oscillations,'' Phys. Rev. {\bf D 22}
(1980) 2718.

\bibitem{Kuo:86}
T. K. Kuo and J. Pantaleone, ``Solar-Neutrino Problem and
Three-Neutrino Oscillations,'' Phys. Rev. Lett. 57, (1986)
1805-1808.

\bibitem{Kim:87}
C.W.~Kim and W.K.~Sze, ``Adiabatic resonant oscillations of solar
neutrinos in three generations,'' Phys. Rev. {\bf D 35} (1987)
1404.

\bibitem{Oli:96}
J.C.~D'Olivo and J.A.~Oteo, ``Nonadiabatic three-neutrino
oscillations in matter,'' Phys. Rev. {\bf D 54}, (1996) 1187-1193.

\bibitem{Ohl:99}
T.~Ohlsson and H.~Snellman, ``Three flavor neutrino oscillations
in matter,'' J.Math.Phys. 41 (2000) 2768-2788; Erratum-ibid. 42
(2001) 2345.

\bibitem{Osl:99}
P.~Osland and T.T.Wu,``Solar Mikheyev-Smirnov-Wolfenstein effect
with three generations of neutrinos,'' Phys. Rev. {\bf D 62}
(2000) 013008.

\bibitem{Leh:00}
H.~Lehmann, P.~Osland and T.T.~Wu, ``Mikheyev-Smirnov-Wolfenstein
Effect for Linear Electron Density,'' Commun.Math.Phys. 219 (2001)
77-88.

\bibitem{Moc:00}
I.~Mocioiu and R.~Shrock, ``Matter Effects on Neutrino
Oscillations in Long Baseline Experiments ,'' Phys. Rev. {\bf D
62} (2000) 053017.

\bibitem{Ohls:00}
T.~Ohlsson and H.~Snellman, ``Three flavor neutrino oscillations
in matter,'' J. Math. Phys. 41 (2000) 2768 {\tt hep-ph/9910546}.

\bibitem{Barger:00}
V.~Barger, S.~Geer, R.~Raja, and K.~Whisnant, ``Three flavor
neutrino oscillations in matter,'' Phys. Rev. {\bf D 62} (1987)
013004.

\bibitem{zzxing:00}
Z.Z.~Xing, ``New Formulation of Matter Effects on Neutrino Mixing
and CP Violation,'' Phys. Lett. B487 (2000) 327-333.

\bibitem{Akh:01}
E.Kh.~Akhmedov, M.A.~Tortola, J.W.F.~Valle , ``A simple analytic
three-flavour description of the day-night effect in the solar
neutrino flux,'' JHEP {\bf 0405} (2004) 057.

\bibitem{Ohl:01}
T.~Ohlsson and H.~Snellman, ``Neutrino oscillations with three
flavors in matter of varying density,'' Eur.Phys.J. C20 (2001)
507-515.

\bibitem{Akh:04}
E.Kh.~Akhmedov, R.~Johansson, M.~Lindner, T.~Ohlsson and
T.~Schwetz, ``Series expansions for three-flavor neutrino
oscillation probabilities in matter,'' JHEP 0404 (2004) 078.

\bibitem{Smi:04}
A.Yu.~Smirnov, ``The MSW effect and Matter Effects in Neutrino
Oscillations,'' {\tt hep-ph/0412391}.

\bibitem{Ble:04}
M.~Blennow, T.~Ohlsson, ``Effective Neutrino Mixing and
Oscillations in Dense Matter ,'' Phys. Lett. B609 (2005) 330-338


\bibitem{Akh:04a}
E.Kh.~Akhmedov, P.~Huber, M.~Lindner and T.~Ohlsson, ``T violation
in neutrino oscillations in matter,'' Nucl.Phys. B608 (2001)
394-422.

\bibitem{Per:99}
O.L.G.~Peres, A.Y.~Smirnov, ``Testing the Solar Neutrino
Conversion with Atmospheric Neutrinos,'' Phys. Lett. B456 (1999)
204-213.

\bibitem{Ioan:04}
A.N.~Ioannisian and A.Y.~Smirnov, ``Neutrino oscillations in low
density medium,'' Phys. Rev. Lett. 93 (2004) 241801 {\tt
hep-ph/0404060}.

\bibitem{Ioan:04a}
A.N.~Ioannisian, N.A.~Kazarian, A. Yu. Smirnov and D. Wyler, ``A
precise analytical description of the Earth matter effect on
oscillations of low energy neutrinos,'' Phys. Rev. {\bf D 71}
(2005) 033006.


\bibitem{KamLAND}
The KamLAND Collaboration (K. Eguchi {\it et al.}), ``Evidence for
Reactor Antineutrino Disappearance,''
    Phys. Rev. Lett. {\bf 90} (2003) 021802.

\bibitem{PREM}
A. M. Dziewonski and D. L. Anderson, ``Preliminary Reference Earth
Model,'' Phys. Earth Planet. Interiors 25, 297 (1981).


\end{thebibliography}
\end{document}